\documentclass[useAMS,usenatbib]{mn2e}
\usepackage{graphicx}
\def\kms{km s$^{-1}$}
\def\eq{\begin{equation}}         
\def\eeq{\end{equation}}      

\title[WR~7a: a V~Sagittae or a qWR star?]{WR~7a: a V~Sagittae or a qWR star?\thanks{Based
on observations made at the Pico dos Dias Observatory/LNA}}
\author[A. S. Oliveira et al.]{A. S. Oliveira $^{1}$\thanks{E-mail:
alex@astro.iag.usp.br (ASO); steiner@astro.iag.usp.br (JES); deo@das.inpe.br (DC)},
J. E. Steiner $^{1}$\footnotemark[2] and D. Cieslinski $^{2}$\footnotemark[2]\\
$^{1}$Departamento de Astronomia - IAG, Universidade de S\~ao Paulo, S\~ao Paulo, Brasil\\
$^{2}$Divis\~ao de Astrof{\'\i}sica, Instituto Nacional de Pesquisas Espaciais, S. J. dos Campos,
Brasil}
\begin{document}

\date{Accepted ????. Received ????; in original form ????}

\pagerange{\pageref{firstpage}--\pageref{lastpage}} \pubyear{2003}

\maketitle

\label{firstpage}

\begin{abstract}

The star WR~7a, also known as SPH~2, has a spectrum that 
resembles that of V~Sagittae stars although no O~VI emission has been reported. 
 The Temporal Variance Spectrum  -- TVS -- analysis of our data shows weak but strongly
  variable emission of O~VI lines which is below the noise level in the intensity spectrum.

Contrary to what is seen in V~Sagittae stars, optical photometric 
monitoring shows very little, if any, flickering. We found evidence of 
periodic variability. The most likely photometric period is $P_{phot} = 0.227(\pm 14)$ d, 
while radial velocities suggest a period of $P_{spec} = 0.204(\pm 13)$ d. 
One-day aliases of these periods can not be ruled out. We call attention to similarities 
with HD~45166 and DI~Cru (= WR~46), where multiple periods are present. 
They may be associated to the binary motion 
or to non-radial oscillations.

In contrast to a previous conclusion by \citet{per}, we show 
that WR~7a contains hydrogen. The spectrum of the primary star seems to be 
detectable as the N~V 4604~{\AA} absorption  
line is visible. If so, it means that the wind is 
optically thin in the continuum and that it is likely to be a helium 
main sequence star. 

Given the similarity to HD~45166, we suggests that WR~7a may be a 
qWR -- quasi Wolf-Rayet -- star. Its classification is WN4h/CE in the \citet{smit}
 three dimensional classification system.

\end{abstract}


\section{Introduction}

The star WR~7a was observed by \citet{sch} - SPH -
 as part of a survey to characterize stars previously identified as H$\alpha$
emission sources. \citet{per} classified it is a WR star of  type
WN4, with a strong emission of C~IV 5801~{\AA}. This line, however, is characteristic of
the WC class. WR~7a is situated in an intermediate stage between the WN and the WN/WC stars,
a category of WR stars discussed by \citet{con}. On the basis of the He~II 
Pickering decrement \citet{per} suggested 
that the star has no hydrogen. Those authors determined a
colour excess of $E(B-V)$ between 
1.2 and 2.1, derived from diffuse interstellar bands, and used the
convolution of the spectrum with a $V$ filter profile 
to measures a $V$ magnitude of 15.0. The star was included in the VIIth
Catalogue of Galactic Wolf-Rayet stars \citep{huc} under the
designation WR~7a and we will here adopt this designation instead 
of the name SPH~2 under which it was previously known.

In a search for new stars of the V~Sagittae class \citep{ste1}, we 
selected WR~7a as a candidate because of the similarity of its spectrum 
to that of V617~Sgr, in spite of the absence of the O~VI lines.

The V~Sagittae stars form a group of 4 galactic binaries:
 V~Sge \citep{her,dia1}, WX~Cen \citep{dia2}, 
V617~Sgr \citep{ste2,cie} and DI~Cru (= WR~46) \citep{marche}. They
are characterized by the presence of strong
emission lines of O~VI and N~V. Moreover, He~II 4686~{\AA} 
is at least two times more intense than H$\beta$ + He~II 4859~{\AA}. 
The spectra of the V~Sagittae stars do not show any 
indication of the secondary component. 
The orbital light curves of the members of this class are either low-amplitude
sine waves or are asymmetric with higher amplitude, 
the latter characterized by primary and secondary eclipses with different depths. 
The orbital periods 
range from 5 to 12 h. The V~Sagittae stars are observationally very 
similar to the Compact Binary Supersoft Sources (CBSS), which are  
more abundant in the 
Magellanic Clouds than in the Galaxy.

The CBSS exhibit very soft X-Ray spectra and high luminosities, 
and are interpreted as suffering hydrostatic
hydrogen nuclear burning on the surface of a white dwarf \citep{heu}.
This burning is due to high mass 
transfer rates from the secondary stars in consequence of the fact
that these systems have secondaries more massive than the primaries 
(see \citet{kah}, for a review and references therein).

In the following section we describe our photometric and spectroscopic 
observations of WR~7a and the reductions of the data, while in section 3 we present the 
results of the analysis. 
The conclusions of this work are presented in section 4. 

\section{Observations}

We made photometric measurements of the star WR~7a with the Boller \& Chivens 60-cm
telescope of the University of S\~ao Paulo and the Zeiss 60-cm telescope,
both located at the Laborat\'orio Nacional de Astrof\'{\i}sica (LNA), in Itajub\'a, 
southeast Brazil. We used the thin, back-illuminated EEV CCD 002-06 chip 
and a Wright Instruments thermoelectrically cooled camera. The timing was 
provided by a Global Positioning System (GPS) receiver. Table~\ref{jour} 
contains a journal of the CCD observations. The images were
obtained through the Johnson $V$ band and were corrected 
for bias and flatfield, using the standard IRAF\footnote{IRAF is distributed 
by the National Optical Astronomy Observatories, which are operated by 
the Association of Universities for Research in Astronomy, Inc., under 
cooperative agreement with the National Science Foundation.} routines.
The star is located in a quite rich field; therefore, we carried out 
differential PSF (Point Spread Function) photometry of WR~7a and four comparison stars
using the LCURVE package, 
written and kindly provided by M.P. Diaz. It makes use of DAOPHOT routines to treat automatically long 
time series data.

\begin{table*}
 \centering
 \begin{minipage}{140mm}
  \caption{Journal of photometric and spectroscopic observations of WR~7a.\label{jour}}
  \begin{tabular}{@{}lllllllll@{}}
  \hline
   Date     & Telescope   & Instrument & CCD & Number of  & Exp. time & Filter & grating  & Resol.\\
   	&	&		&	&exps.	&(s)	&	&(l/mm)&	({\AA})\\
   \hline
23/01/2000	& B\&C 60 cm	& Imaging Camera 	&301&165      	&150	&V	&...&...\\
02/02/2000	& Zeiss 60 cm	& Imaging Camera 	 &301&1180	&15	&V	&...&... \\
02/05/2001	& Zeiss 60 cm	& Imaging Camera 	 &301&363	&20	&V	&...&... \\
03/05/2001	& Zeiss 60 cm	&Imaging Camera 	 &301&360	&30	&V	&...&... \\
02/02/2000	& B\&C 1,6 m	& Cassegrain Spectr. &106&32	&600	&...&900	&2.8	\\
03/02/2000	& B\&C 1,6 m	&Cassegrain Spectr.  &106&11	&900	&...&900	&2.8	\\
\hline
\end{tabular}
\end{minipage}
\end{table*}

Spectroscopic measurements of WR~7a were carried out with the Cassegrain
spectrograph coupled to the 1.6-m telescope at the LNA.
We used a 900 lines per mm dispersion grating, covering the spectrum from 
3650~{\AA} to 4950~{\AA} --  see Table~\ref{jour}. 
We used a thin, back-illuminated SITe SI003AB 1024x1024 CCD.
Bias and dome flatfield exposures were obtained 
to correct for the read-out pattern and 
sensitivity of the CCD. The width of the slit was adjusted to the seeing
conditions at the time of observations. We took exposures of calibration 
lamps after every third exposure of the star, in order to determine the
wavelength calibration solution. The image reductions, spectra extractions 
and wavelength calibrations were executed with IRAF standard routines,
and the measurements of the He~II 4686~{\AA} radial velocities were done by fitting
gaussian functions to the peaks of these emission lines.

\section{Data analysis and discussion}

\subsection{Analysis of the optical spectrum: the variable O~VI lines}

\begin{figure}
\vspace*{15pt}
\centerline{\includegraphics[width=84mm]{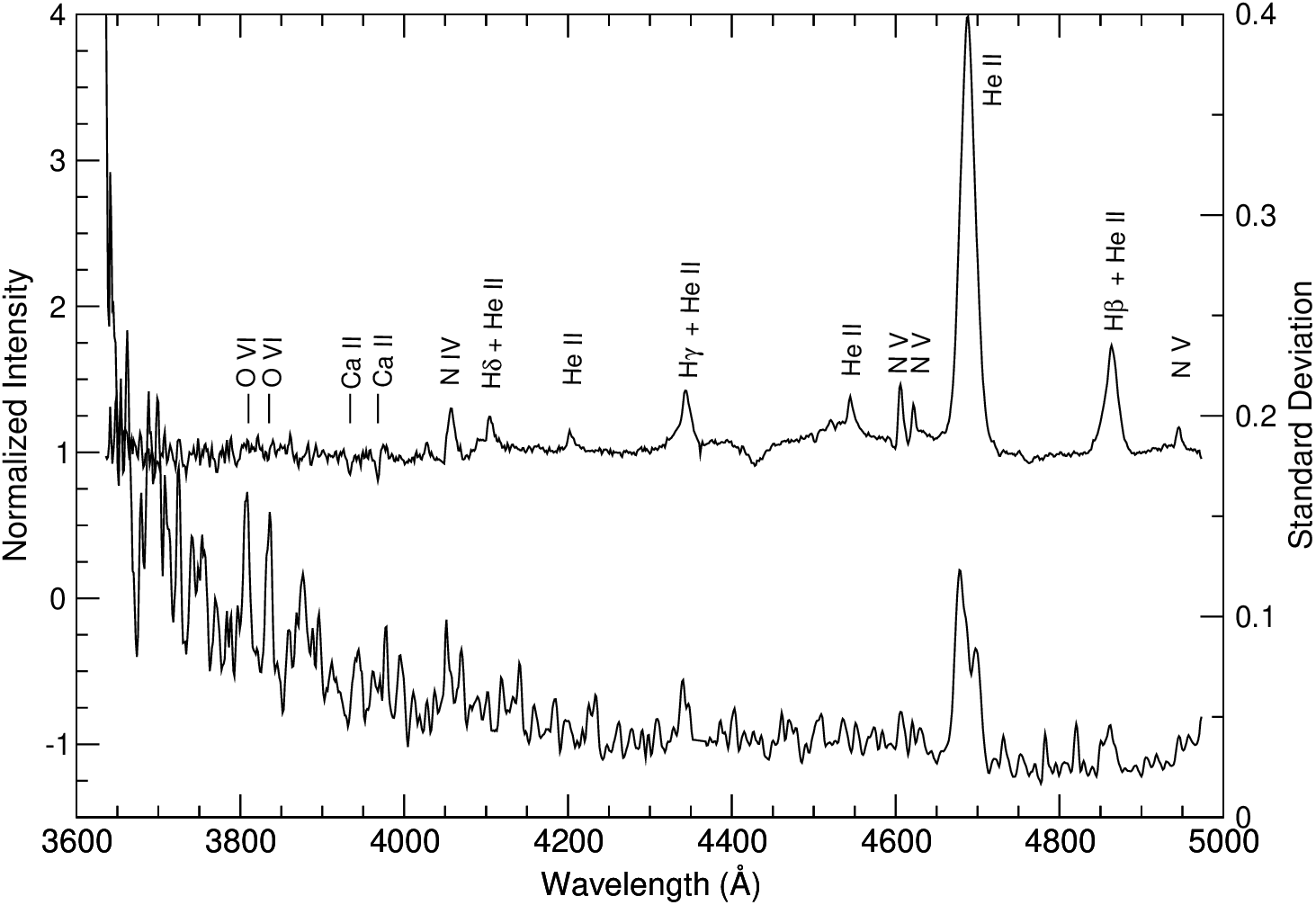}}
\caption{Intensity spectrum (above) and Time Variance Spectrum (below) 
of WR~7a. \label{desv}}
\end{figure}

The spectrum of WR~7a is very similar to those of V~Sagittae stars, except for 
the absence of O~VI emission lines. In particular, the similarity 
between its spectrum and that of V617~Sgr 
\citep{ste2} is remarkable. The properties of the most prominent lines of WR~7a are 
presented in Table~\ref{lines}.  In order to examine in more detail 
the characteristics of the spectrum we performed a Temporal Variance Spectrum 
(TVS) analysis. In this analysis we normalize all spectra in such a way that 
the continuum is equal to 1. The temporal variance is calculated for each wavelength 
pixel. In our TVS spectra we calculated  the square root of the variance as a function 
of wavelength. For a more detailed discussion of this method see \citet{fulle}.
 Figure~\ref{desv} shows the 
the intensity spectrum and the TVS for WR~7a. Although below the noise
level in the intensity spectrum, the O~VI  3811/34~{\AA} emission lines are easily visible in 
the TVS, indicating their high variability. WR~7a displays, therefore, all 
the spectroscopic features that characterize the V~Sagittae stars such as 
the high He~II/H$\beta$ ratio and the simultaneous 
presence of O~VI and N~V lines.

\begin{table*}
 \centering
 \begin{minipage}{140mm}
  \caption{Spectral line properties of WR~7a.\label{lines}}
  \begin{tabular}{@{}llll@{}}
  \hline
  Species      & Observed		& W$_{\lambda}$ & FWHM \\
   	     &wavelength ({\AA})	& ({\AA})             & (\kms)  \\
   \hline
Ca II		& 3933.73	&0.54	& 249:	\\
Ca II		& 3967.77	&0.57	& 147:	\\
He II		&4028.00	&-0.75	&420	\\
N IV		&4057.70	&-4.2	&763	\\
H$\delta$	&4104.93	&-1.0	&436	\\
He II		& 4203.5:	                 &-1.5:	& 884:	\\
H$\gamma$	& 4343.7:	                 &-5.6:	& 1109:	\\
He II		& 4544.60	&-0.45	& 206   	\\
N V		& 4604.24	&-2.9	& 383  	\\
N V		& 4621.82	&-1.2	& 276   	\\
He II		& 4688.10	&-63	& 1389	\\
H$\beta$		& 4863.93	&-15.3	& 1209	\\
N V		& 4945.51	&-0.90	&349	\\
\hline
\end{tabular}
\end{minipage}
\end{table*}

\subsection{Searching for periodicities}

The light curves of the four observed nights are quite flat, with very 
little, if any, flickering. V~Sagittae stars have, as a rule, significant 
variability on time-scales of tens of minutes. In addition, the 
orbital variations are easily recognizable in light curves of V~Sagittae
stars but not in WR~7a. Thus,
from the photometric point of view, WR~7a does not behave like 
the V~Sagittae stars. 

\begin{figure}
\vspace*{15pt}
\centerline{\includegraphics[width=84mm]{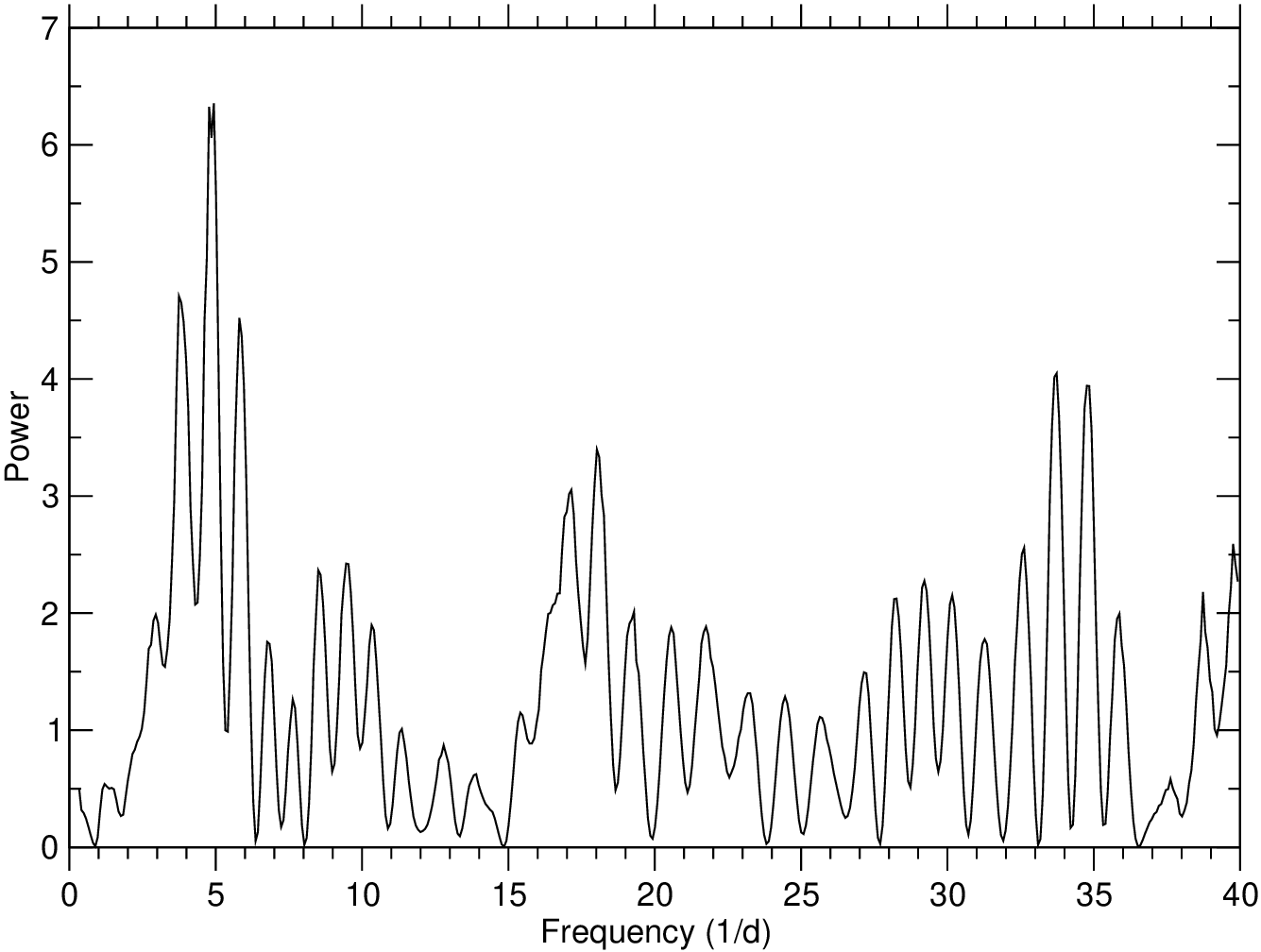}}
\caption{Lomb-Scargle periodogram of the He~II 4686~{\AA} radial velocities data. \label{lombvr}}
\end{figure}

\begin{figure}
\vspace*{15pt}
\centerline{\includegraphics[width=84mm]{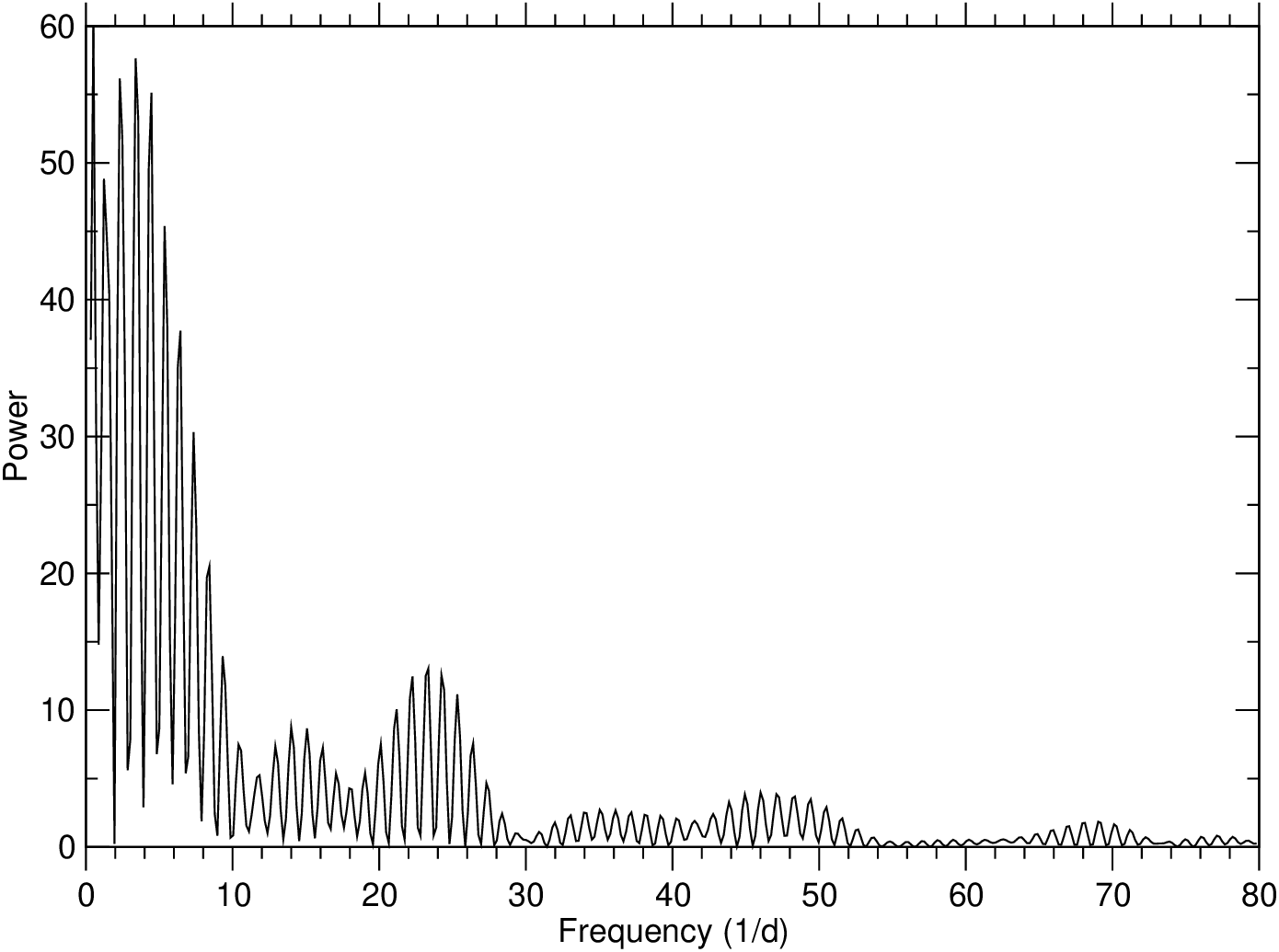}}
\caption{Lomb-Scargle periodogram of the photometry data. \label{lombfot}}
\end{figure}

We used our radial velocity and 
photometric data to search for possible periodicities. The Lomb-Scargle 
periodogram \citep{sca} of the He~II 4686~{\AA} radial velocity measurements 
from our spectra is presented in Figure~\ref{lombvr}. 
We also show, in Figure~\ref{lombfot}, the periodogram of the photometric 
measurements. The latter was constructed on the basis of data from
the two consecutive nights of 2000 May 2 and 3. The data from the other observing
nights were non-consecutive and too far apart to be included in 
the periodogram analysis. 

The highest peak seen in the 
Lomb-Scargle periodogram (Figure~\ref{lombvr}) of the radial velocity 
data corresponds to a period of 0.204 d ($f=4.9 ~$d$^{-1}$). The
photometric data (Figure~\ref{lombfot}) yield a slightly different value, 
0.227 d ($f=4.4 ~$d$^{-1}$). Given the limited time coverage of 
our data, the errors involved are quite large.
 The two periods differ by $1.7\sigma$. This means that there is a non-negligible
 chance that they are actually identical. One-day aliases of these periods
  can not be ruled out. Period values such as these,
or multiple periodicities, are not unusual in systems like WR~7a.
The orbital period of the V~Sagittae star V617~Sgr, for instance, is 0.208 d \citep{ste2},
while the qWR star HD~45166 (Steiner \& Oliveira, in preparation) and the WR star
DI~Cru \citep{marche, vee2}
are systems with multiple periods that probably include the 
orbital period plus a number of possible periods due to non-radial 
pulsations. 
We do not claim that WR~7a is necessarily a binary system. 
Candidates to non-radial pulsations in the stars HD~45166 and DI~Cru are
 in the range from 0.2 to 0.4 d. 
 Given the experience with these systems, a lot more spectra with 
higher spectral resolution and higher signal-to-noise ratio are required to clarify 
the nature of these periods -- or, eventually, of this single period -- in WR~7a.

\begin{figure}
\vspace*{15pt}
\centerline{\includegraphics[width=84mm]{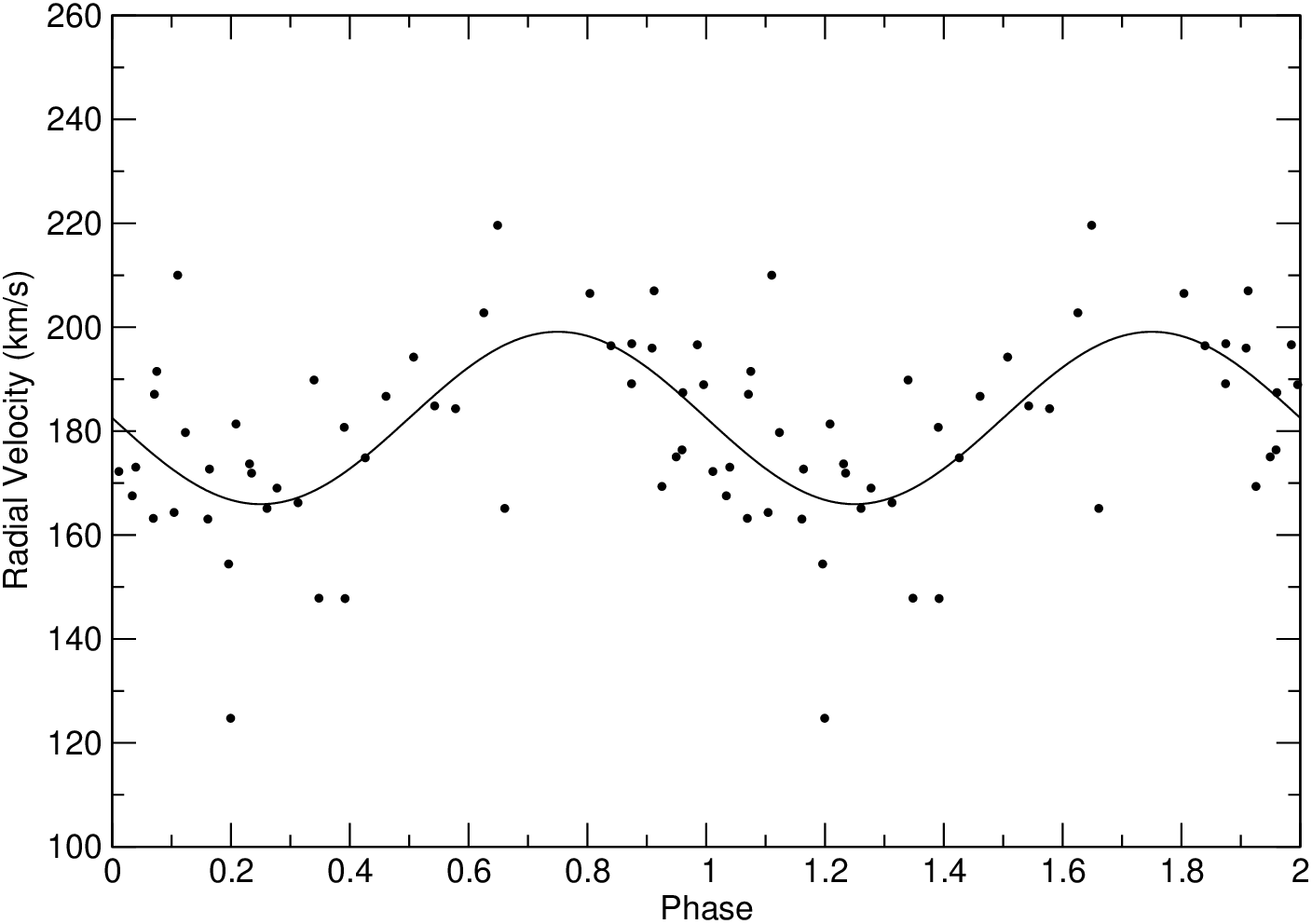}}
\caption{Radial velocity curve of the He~II 4686~{\AA} emission line, 
folded to the 0.204 d period. The continuous
line is the best-fitting sine curve. \label{vrcurv}}
\end{figure}

\begin{figure}
\vspace*{15pt}
\centerline{\includegraphics[width=84mm]{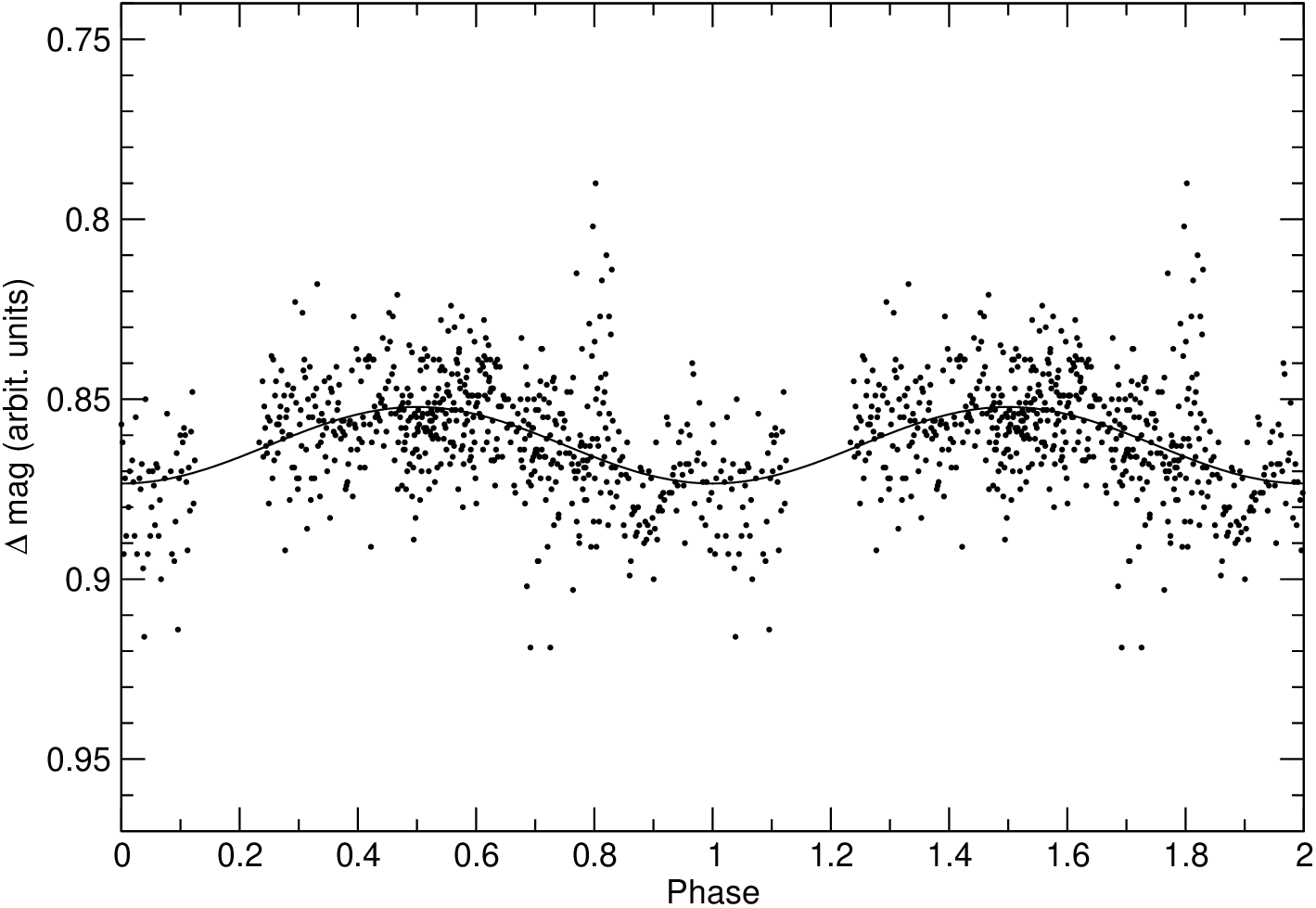}}
\caption{Average light curve folded to the 0.227 d period. The continuous
line is the best-fitting sine curve. \label{lccurv}}
\end{figure}

The radial velocity and photometric ephemerides are

\eq
T_{spec} (HJD) = 2~451~577.692(\pm 20) + 0.204(\pm 13) \times E
\eeq
and
\eq
T_{phot} (HJD) = 2~452~032.600(\pm 23) + 0.227(\pm 14) \times E
\eeq
where zero phase of the radial velocity ephemeris is defined as the crossing from positive to negative 
values, and $T_{phot}$ are timings of minimum brightness. 
The radial velocity curve of the He~II 4686~{\AA} line, folded to 
the 0.204 d period is shown in Figure~\ref{vrcurv}. It presents 
sinusoidal variations with an amplitude of about $K \sim~17$ \kms.
The average light curve folded to the 0.227 d period is shown in Figure~\ref{lccurv},
and presents a peak-to-peak amplitude of 0.02 mag.

An additional (weak) signal is seen in the photometric periodogram 
(Figure~\ref{lombfot}), where one can see power excess at about 60 
min ($f=24~$d$^{-1}$). Theoretical studies on radial pulsations 
of Wolf-Rayet stars \citep{maed} indicate oscillation periods in the 
range of 15 to 60 min. \cite{blecha} claim to have seen
similar oscillations with period of 10 min in WR~40,
but their reality has been questioned \citep{marche2}. 
Further investigation 
of this particular issue is certainly desirable.

\subsection{The Hydrogen content}

\begin{figure}
\vspace*{15pt}
\centerline{\includegraphics[width=84mm]{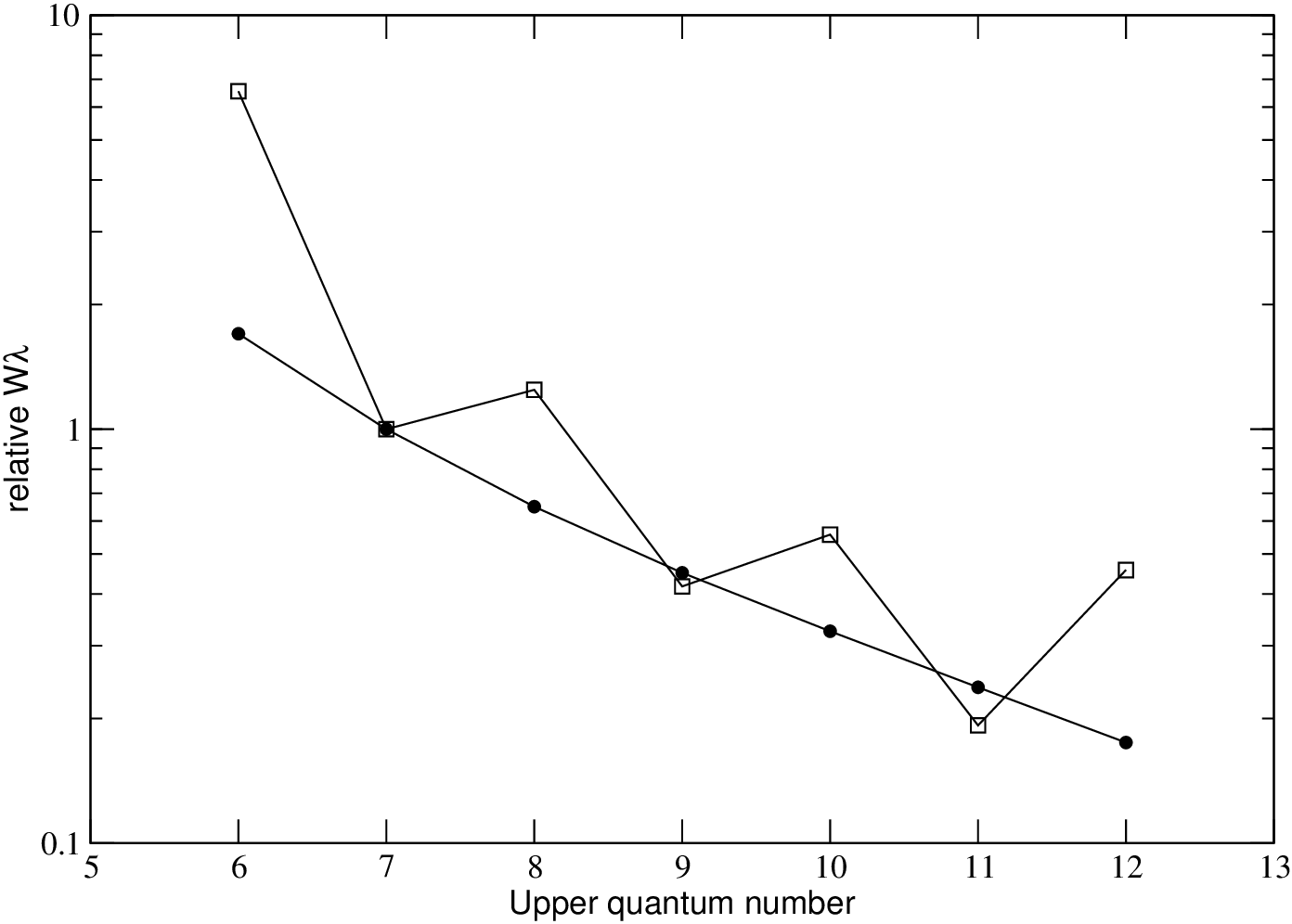}}
\caption{The He~II Pickering series for WR~7a. Theoretical intensities for a pure He~II spectrum are 
represented by filled circles and observed data are represented by empty 
squares.  \label{pick}}
\end{figure}

\citet{per} claimed that WR~7a does not present hydrogen 
emission. They concluded this on the basis of their He~II Pickering decrement 
analysis. A re-analysis of the equivalent widths presented by these authors, 
however, shows that this conclusion is not correct. We applied a procedure in 
which we compared theoretical \citep {ost} and observed equivalent 
widths of the He~II lines of the Pickering series, normalized to the 
equivalent width of
He~II~5411~{\AA}. The result is shown in Figure~\ref{pick}. 
The oscillatory behavior associated to the presence of hydrogen is 
clearly noticed.

Another way to investigate the presence/absence of hydrogen in a spectrum 
with strong helium emission consists in comparing the He~II 4859~{\AA}+H$\beta$ strength
with the geometric mean of the strength of the two 
adjacent transitions from the Pickering series. This 
criterion has been widely used, for instance, by \citet{smit} in defining 
their three-dimensional classification of WN stars. We define the Pickering 
parameter, $p$, such that

\eq
p = \frac{I(4859\mathrm{\AA}+4861\mathrm{\AA})}{\left(I(4541\mathrm{\AA}) 
\times I(5411\mathrm{\AA})\right)^{1/2}}
\eeq

For a pure He~II spectrum one expects $p=1$ \citep{smit}. Any value of $p$ 
larger than 1 would mean that hydrogen is present in the spectrum. In 
the case of WR~7a, the measured value is about $p = 1.88$.

If both hydrogen and helium 
lines were optically thin, it would be easy to determine the relative 
abundance between the two species. In this case \citep{cont},

\eq
\frac{\mathrm{N}(H^{+})}{\mathrm{N}(He^{++})} = p - 1
\eeq
and H/He $\sim~ 0.88$. The largest uncertainty 
in this determination comes from the hypothesis that all of the involved 
lines are optically thin.

For the optically thick case, one gets (see \citet{cont})

\eq
\frac{\mathrm{N}(H^{+})}{\mathrm{N}(He^{++})} = p^{3/2} - 1
\eeq
and the abundance is H/He $\sim ~1.6$. As the 
TVS analysis shows some absorption in the He~II line,
 it is probably optically thick at low velocities. At high velocities the wind is usually
 optically thin. In the present case we have, therefore, a situation that is likely
 to be intermediate between optically thick (line core) and optically thin (line wings)
 and the actual abundance is likely to be intermediate between the two values mentioned above.

 \subsection{Do we see a photospheric absorption line? -- Comparison with HD~45166}

In the spectrum of WR~7a published by \citet{per} one can see a weak, but clearly
present, absorption line at 4604{\AA}. It 
is also (albeit barely) visible in our spectrum of poorer 
signal-to-noise ratio. This is not likely to be a P Cyg absorption 
because such features are not seen in the Balmer or He~II 
lines. 

We have seen a similar feature in the high resolution and high signal-to-noise  
spectrum of the qWR star HD~45166 (Steiner \& Oliveira, in 
preparation). In that case, we identified this feature as a N~V photospheric line 
of the hot primary, which is a helium main sequence star of about $3.5~M_{\sun}$.
The IUE spectrum of HD~45166 \citep{willi2} also shows photospheric lines of N~V,
Fe~V, O~V among others. This is a clear evidence that, in that star, the wind is optically
thin at the continuum, contrary to what is usually the case in WR stars.
This is not surprising, given that the mass-loss rate is about 1000 times smaller in HD~45166 
than in WR stars. As the mass-loss rate is related to the stellar luminosity, and the latter to its mass
(see \citet{lamer}), and since we identify the N~V absorption line of WR~7a as
a photospheric feature, we conclude that WR~7a has a mass which is smaller than 
the mass of a typical WR star. On the other hand, the mass cannot be too small, otherwise 
its temperature would not be high enough to produce the N~V lines or the O~VI lines.
The emission lines in WR~7a have equivalent widths that are about a factor of two
stronger than in HD~45166, after correcting, in the latter case, for the strong contribution
of the secondary. This also suggests that the mass-loss rate and the luminosity -- and consequently,
the mass -- is higher than in HD~45166. In conclusion, WR~7a could be interpreted 
as a helium main sequence object with a mass that is intermediate between HD~45166
($3.5 M_{\sun}$) and typical WR stars ($\sim 10 M_{\sun}$).

\subsection{How to classify WR~7a?}

The spectral characteristics of WR~7a indicate that it could be classified as a 
star of the V~Sagittae type. It presents all the characteristics that define this class, 
such as the intensity ratio of He~II 4686~{\AA} and H$\beta$ as well as the simultaneous 
presence of N~V 4945~{\AA} and O~VI 3811/34~{\AA} emission lines. 
However, the presence of the N~V 4604~{\AA} photospheric 
absorption line and the absence of flickering argue against this classification and 
point towards the similarities to HD~45166 (Steiner \& Oliveira, in preparation), which has 
been classified as a qWR star (see \citet{willi} and references therein). 
Although the observational differences between these two groups of stars are 
quite subtle, they follow very distinct paradigms in terms of their ultimate nature. 
While V~Sagittae stars are interpreted as the galactic counterpart of CBSS and, 
therefore, accreting white dwarfs with hydrogen burning on their surface \citep{ste1,heu}, the 
qWR paradigm assumes a 
helium main sequence star (Steiner \& Oliveira, in preparation).  

In the three-dimensional classification system proposed by \citet{smit} for WN 
stars, we classify WR~7a as WN4h/CE. This takes into account the degree of 
ionization (He~II to He I line ratio), the presence and abundance of hydrogen, the He~II 4686~{\AA}
line width and 
the strength of the C IV 5801/11~{\AA} emission lines.

\section{Conclusions}

	The main conclusions of this paper are:

\begin{enumerate}
\renewcommand{\theenumi}{(\arabic{enumi})}

\item The star WR~7a = SPH~2 has a spectrum that resembles that of V~Sagittae 
stars.
\item TVS analysis shows strong variability in the O~VI lines. They are below the noise level 
in the intensity spectrum.
\item Contrary to what is seen in V~Sagittae stars, optical photometric monitoring 
shows very little, if any, flickering.
\item We found evidence of periodic variability. The most likely 
spectroscopic period is $P_{spec}=0.204$ d, with $K \sim 17$ \kms.
Photometry suggests a period of $P_{phot}=0.227$ d,
and a peak-to-peak amplitude of 0.02 mag. One-day aliases of these periods can not be ruled out. 
We do not claim that these periods are distinct, nor that WR~7a is necessarily a binary system. 
\item We call attention to 
similarities with HD~45166 and DI~Cru, where multiple periods are 
present, which may be associated to the binary motion and/or non-radial
 oscillations.
\item Contrary to what was claimed by \citet{per}, we show 
that WR~7a contains hydrogen. We determine the H/He abundance, by number, to be
between 0.88 (optically thin approximation) and 1.6 (optically thick approximation).  
\item The spectrum of the primary star seems to be detectable as the N~V 
4604~{\AA} absorption line is visible. If so, it means that the wind is optically thin in the 
continuum and that WR~7a is likely to be a helium main sequence star.
\item Given the similarity to HD~45166, we suggest that WR~7a may be a 
qWR (quasi Wolf-Rayet) star. Its WR classification is WN4h/CE. 

\end{enumerate}

\section{Acknowledgments}
 
 We thank Dr. Albert Bruch for a critical reading of the manuscript.

\label{lastpage}

\end{document}